\documentclass[aps,twocolumn,showpacs,pre]{revtex4}
\newcommand{\n}{\noindent}
\begin{document}

\title{Quantum dynamical echo in two-level systems} 
\author{R. Sankaranarayanan, Jane H. Sheeba and M. Lakshmanan}
\affiliation{Centre for Nonlinear Dynamics, Department of Physics \\
Bharathidasan University, Tiruchirappalli 620024, India.}

\begin{abstract}
Evolution of quantum fidelity for two-level systems is studied in the
context of periodic echo. From a general treatment for time independent
case, we obtain a simple condition on the governing Hamiltonians under
which the systems display periodic quantum echo. For a specific time
dependent problem the quantum fidelity is shown to exhibit Rabi
oscillation. This may be considered as a simple mechanism to generate
periodic echo, except for a specific initial superpositional state in
which case the fidelity remains invariant.
\end{abstract}
\maketitle

\section{Introduction}

Since the overlap of states is invariant under unitary quantum dynamics,
a new approach is necessary for measuring sensitivity of the evolution with
respect to perturbation. A seminal idea for such investigation is to consider
two evolutions of a single initial state corresponding to two slightly
different Hamiltonians, as proposed by Peres \cite{peres84}. Let us consider
a quantum system, which is initially in an arbitrary state $|\alpha\rangle$,
evolves under the Hamiltonians $H$ and $\tilde H = H + \epsilon V$ with
$\epsilon$ being perturbation parameter. If $U(t)$ and $\tilde U(t)$ are the
corresponding evolution operators, the overlap intensity of the two evolved
states or quantum fidelity 
\begin{equation}
f(t) = |\langle\alpha|\tilde U^{\dagger}(t)U(t)|\alpha\rangle|^2
\label{fid}
\end{equation}

\n is a measure of quantum sensitivity. This quantity can also be interpreted
as the overlap between an evolved state, which is obtained by forward evolving
in time under $H$ followed by backward evolution for the same time $t$ under
$\tilde H$, and the initial state. Latter interpretation suggests that quantum
fidelity (also called as Loschmidt echo) measures the accuracy to which a
quantum state can be recovered by inverting the perturbed evolution.

Only in recent years, Peres' proposal has evoked enormous interest with
different interpretations. In particular, exponential decay of $f(t)$ in
time has been identified as quantum manifestation of classically chaotic
systems \cite{jala01}. Since fidelity is intimately related to decoherence
\cite{cucc03}, the decaying nature could be a concern for practical
implementation of algorithms to perform quantum computation. Loschmidt echo
is realized as spin (polarization) echo in nuclear magnetic resonance
experiments \cite{zhang92}, and attenuation of echo amplitude is found
to be useful tool in characterizing many-body quantum systems \cite{usa98}.
Fidelity is also a relevant quantity in spectroscopic studies on quantum
stability of optically trapped two-level atoms \cite{ander03}. 

While for most of the dynamical systems the quantum fidelity is either
a decaying function or exhibits attenuated oscillations in time, for
the harmonic oscillator it is periodic in first order perturbative
approximation \cite{sankar03}. We may then enquire about other physical
situations wherein the fidelity can be periodic. If $f(t)=f(t+jT)$ with
integer $j$, $f(jT)=1$ as $f(0)=1$. That is to say that, after reversing
the perturbed dynamics at $t=jT$ the initial state $|\alpha\rangle$ is
completely recovered (up to the phase factor). We refer to this as
{\it periodic quantum echo} with period $T$. Before addressing the
possibility and utilization of such echo in many-body complex systems,
it is instructive to resolve the dynamics of two-level quantum system (qubit)
in this context. It goes without saying that any understanding of qubit is
of fundamental importance in the subject of quantum computation \cite{niel00}
as well. In the present note, we compute quantum fidelity to explore the echo
scenario for two-level systems.

It should be noted that $f(t)$ depends on the form of perturbation and
the choice of initial state. If the initial state is an eigenstate of one
of the Hamiltonians (time independent), quantum fidelity is nothing but
the survival probability under the evolution governed by the other
Hamiltonian. Further, for a two-level system the survival probability is
a simple periodic function in time. Hence we naively expect the fidelity
to be a non-decaying function, for an arbitrary choice of initial state.
In the next section, we show for time independent case that in general
the quantum fidelity is quasiperiodic. It will be periodic provided the
ratio of eigen-energy spacings of the two Hamiltonians are rational,
implying that the two-level systems exhibit periodic quantum echoes.
In section III, we dwell on such echo scenario for a specific time dependent
problem. For a time periodic perturbation, it is shown that the well known
Rabi oscillation of two-level system is directly reflected on the temporal
behaviour of quantum fidelity. While this is seen as a simple mechanism
to observe periodic quantum echoes in two-level systems, we point out that
for certain initial superpositional states the Rabi oscillation vanishes
and the fidelity is invariant in time.

\section{Time independent case}

Let us first consider two time independent Hamiltonians $H$ and $\tilde H$,
whose eigenvalue equations are 
\begin{equation} 
H|\phi_n\rangle = E_n|\phi_n\rangle \; ; \; 
\tilde H|\tilde{\phi}_m\rangle = \tilde{E}_m|\tilde{\phi}_m\rangle \, ,
\label{eigen}
\end{equation}

\n where $m,n=1,2$. Note that the eigen sets $\{|\phi_n\rangle\}$ and
$\{|\tilde\phi_m\rangle\}$ individually form an orthonormal basis. Expanding
the initial state $|\alpha\rangle$ in the eigen basis, one can write
\begin{equation}
|\alpha\rangle = u_1|\phi_1\rangle+u_2|\phi_2\rangle
= v_1|\tilde{\phi}_1\rangle+v_2|\tilde{\phi}_2\rangle \, ,
\end{equation}

\n where $u_1,u_2$ and $v_1,v_2$ are complex numbers satisfying the
normalization conditions, $|u_1|^2 + |u_2|^2 = |v_1|^2 + |v_2|^2 = 1$.
Then the fidelity, defined by eq. (\ref{fid}), can be expressed as
\begin{equation}
f(t) = \left| \sum_{m,n} e^{i(\tilde{E}_m-E_n)t/\hbar}v_m^*u_n
\langle\tilde{\phi}_m|\phi_n\rangle \right|^2 \, .
\end{equation}

\n Introducing the `frequency' variables
\begin{equation}
\omega = {E_2-E_1\over\hbar} \; , \;
\tilde\omega = {\tilde E_2-\tilde E_1\over\hbar}
\end{equation}

\n after some algebra the fidelity can be recasted as 
\begin{widetext}
\begin{eqnarray}
f(t) &=& |v_1|^2 \Big[ |u_1|^2|\langle\tilde{\phi}_1|\phi_1\rangle|^2
+ |u_2|^2|\langle\tilde{\phi}_1|\phi_2\rangle|^2 \Big] 
+ |v_2|^2 \Big[ |u_1|^2|\langle\tilde{\phi}_2|\phi_1\rangle|^2
+ |u_2|^2|\langle\tilde{\phi}_2|\phi_2\rangle|^2 \Big] \nonumber \\ 
&& + \; 2\;\Bigg\{ \Big(|v_1|^2-|v_2|^2\Big)
\Big[a_1\cos\omega t-b_1\sin\omega t\Big]
+ \Big(|u_1|^2-|u_2|^2\Big) \Big[a_2\cos\tilde\omega t
- b_2\sin\tilde\omega t\Big] \nonumber\\
&& \;\;\;\;\;\;\;\;\; + \Big[a_3\cos(\tilde\omega-\omega)t
+ b_3\sin(\tilde\omega-\omega)t\Big] + \Big[a_4\cos(\tilde\omega+\omega)t 
+ b_4\sin(\tilde\omega+\omega)t\Big] \Bigg\} \, ,
\label{fid1}
\end{eqnarray}
\end{widetext}
\n where
\[ \begin{array}{lll}
a_1+ib_1 &=& u_1u_2^*\langle\tilde{\phi}_1|\phi_1\rangle
\langle\phi_2|\tilde{\phi}_1\rangle \, , \\[4pt] 
a_2+ib_2 &=& v_1v_2^*\langle\tilde{\phi}_2|\phi_1\rangle
\langle\phi_1|\tilde{\phi}_1\rangle \, , \\[4pt] 
a_3+ib_3 &=& u_1u_2^*v_1^*v_2\langle\tilde{\phi}_1|\phi_1\rangle
\langle\phi_2|\tilde{\phi}_2\rangle \, , \\[4pt]
a_4+ib_4 &=& u_1^*u_2v_1^*v_2\langle\tilde{\phi}_1|\phi_2\rangle
\langle\phi_1|\tilde{\phi}_2\rangle \, , \\ 
\end{array} \]

\n with $a_i$'s and $b_i$'s being real constants. In this form, the
fidelity is a sum of four periodic functions and hence, in general $f(t)$
is quasiperiodic. We may then look at the expression (\ref{fid1}) for the
following choices of initial states. \\

\n (i) $|\alpha\rangle = |\phi_1\rangle$; $u_1=1,u_2=0$:
\begin{eqnarray}
f(t) &=& {|\langle\tilde\phi_1|\phi_1\rangle|}^4
+ {|\langle\tilde\phi_2|\phi_1\rangle|}^4 \nonumber \\
&& \hspace*{1cm} + \;2\; {|\langle\tilde\phi_1|\phi_1\rangle|}^2 
{|\langle\tilde\phi_2|\phi_1\rangle|}^2 \cos\tilde{\omega}t \, .
\end{eqnarray}

\n (ii) $|\alpha\rangle = |\phi_2\rangle$; $u_1=0,u_2=1$: 
\begin{eqnarray}
f(t) &=& {|\langle\tilde\phi_1|\phi_2\rangle|}^4
+ {|\langle\tilde\phi_2|\phi_2\rangle|}^4 \nonumber \\
&& \hspace*{1cm} + \;2\; {|\langle\tilde\phi_1|\phi_2\rangle|}^2 
{|\langle\tilde\phi_2|\phi_2\rangle|}^2 \cos\tilde\omega t \, .
\end{eqnarray}

\n Since the $H$-evolution introduces just a phase for the above two
choices of initial state, fidelity is nothing but the survival probability
in $\tilde H$-evolution. It is a simple periodic function in time with
period $T = 2\pi/|\tilde\omega|$. \\ 

\n (iii) $|\alpha\rangle = |\tilde\phi_1\rangle$; $v_1=1,v_2=0$:
\begin{eqnarray}
f(t) &=& {|\langle\phi_1|\tilde\phi_1\rangle|}^4
+ {|\langle\phi_2|\tilde\phi_1\rangle|}^4 \nonumber \\
&& \hspace*{1cm} + \;2\; {|\langle\phi_1|\tilde\phi_1\rangle|}^2 
{|\langle\phi_2|\tilde\phi_1\rangle|}^2 \cos\omega t \, .
\end{eqnarray}

\n (iv) $|\alpha\rangle = |\tilde\phi_2\rangle$; $v_1=0,v_2=1$:
\begin{eqnarray}
f(t) &=& {|\langle\phi_1|\tilde\phi_2\rangle|}^4
+ {|\langle\phi_2|\tilde\phi_2\rangle|}^4 \nonumber \\
&& \hspace*{1cm} + \;2\; {|\langle\phi_1|\tilde\phi_2\rangle|}^2 
{|\langle\phi_2|\tilde\phi_2\rangle|}^2 \cos\omega t \, .
\end{eqnarray}

\n For the last two cases, $\tilde H$-evolution introduces just a phase.
Hence, $f(t)$ is the survival probability in $H$-evolution and it oscillates
with period $T = 2\pi/|\omega|$. 

On the other hand, for an arbitrary choice of initial state the fidelity is
the sum of four periodic functions each of them oscillating with frequencies
$\omega,\tilde\omega,(\tilde\omega-\omega)$ and $(\tilde\omega+\omega)$.
Then $f(t)$ can be periodic if and only if all the four frequencies are
commensurate with each other. Since the last two frequencies are just the
sum and difference of the first two, the required condition for $f(t)$ to
be periodic is
\begin{equation}
{\tilde\omega\over\omega} = {\tilde E_2-\tilde E_1\over E_2-E_1}
= {p\over q} \, ,
\label{rational}
\end{equation} 

\n where $p$ and $q$ are co-primes. Representing the Hamiltonian $H$
in a specific two-level basis states $|1\rangle$ and $|2\rangle$, energy
eigenvalues are solutions of the quadratic equation
\begin{equation}\left|
\begin{array}{cc} H_{11}-E & H_{12} \\ H_{12}^{*} & H_{22}-E \end{array}
\right| = 0 \, ,
\end{equation}

\n where $H_{ij}$ are matrix elements of $H$ in the chosen basis. Denoting
the two solutions as $E_1$ and $E_2$, we have 
\begin{equation}
E_2 - E_1 = \sqrt{(H_{11}-H_{22})^2+4|H_{12}|^2}  \, .
\end{equation}

\n With similar expression for the Hamiltonian $\tilde H$, the condition for
periodic fidelity becomes
\begin{eqnarray}
&& \left({p\over q}\right)^2 \Big[(H_{11}-H_{22})^2+4|H_{12}|^2 \Big]
\nonumber \\ && \hspace*{2cm} = (\tilde{H}_{11}-\tilde{H}_{22})^2 
+ 4|\tilde{H}_{12}|^2 \, .
\label{HH}
\end{eqnarray}

\n Thus we arrive at a condition on matrix elements of the two Hamiltonians
such that the quantum fidelity of an arbitrary state is periodic and
the period $T$ can be calculated as follows. If $\tilde\omega=\omega$, 
\begin{equation}
T=2\pi\hbar/|E_2-E_1| \, . 
\end{equation}

\n For $\tilde\omega\neq\omega$, we have a relation between the frequencies
deduced from eq. (\ref{rational}) as 
\begin{eqnarray}
p(p^2-q^2)\omega &=& q(p^2-q^2)\tilde\omega 
= pq(p+q)(\tilde\omega-\omega) \nonumber \\ 
&& \hspace*{1cm} = pq(p-q)(\tilde\omega+\omega) \, ,
\end{eqnarray} 

\n implying that 
\begin{equation}
T = 2\pi\hbar/|p(p^2-q^2)(E_2-E_1)| \, .
\end{equation}

\n With this, we summarize our general treatment on time independent problem.
For a given forward evolution of an arbitrary initial state $|\alpha\rangle$
governed by the Hamiltonian $H$, we shall choose another Hamiltonian
$\tilde H$ such that its matrix elements in the chosen basis satisfies the
condition (\ref{HH}). Then by reversing the evolution with $\tilde H$ at
$t=jT$ ($j$ is integer), the initial state is completely recovered. We have
then the scenario of periodic quantum echo.

\subsection{Perturbative approximation}

If we consider $\tilde H$ as a perturbed Hamiltonian, that is, 
$\tilde H = H +\epsilon V$ with $|\epsilon| \ll 1$, we shall use the
first order time-independent expansions for the eigenvalues 
\begin{equation}
\tilde E_1 \approx E_1 + \epsilon V_{11} \;\;\; ; \;\;\; 
\tilde E_2 \approx E_2 + \epsilon V_{22} \, ,
\label{eval}
\end{equation}

\n and for the eigenfunctions 
\begin{widetext}
\begin{equation}
|\tilde\phi_1\rangle \approx |\phi_1\rangle 
- {\epsilon V_{21}\over (E_2-E_1)} |\phi_2\rangle \;\;\; ; \;\;\; 
|\tilde\phi_2\rangle \approx |\phi_2\rangle 
- {\epsilon V_{12}\over (E_1-E_2)} |\phi_1\rangle \, ,
\end{equation}

\n where $V_{mn} = \langle\phi_m|V|\phi_n\rangle$. Note that the above
expansions are valid for $|\epsilon V_{12}| \ll |E_1-E_2|$. Then the
expression (\ref{fid1}) can be approximated up to the order in $\epsilon$ as 
\begin{eqnarray}
f(t) &\approx& |u_1|^4 + |u_2|^4 + 2|u_1|^2|u_2|^2\cos\omega_\epsilon t
-2\epsilon\;{|u_1|^2-|u_2|^2\over E_2-E_1}\;\Bigg\{ A + \Big[A\cos\omega t
+ B \sin\omega t\Big] \nonumber \\
&& \hspace*{5cm} 
- \Big[A\cos\omega_\epsilon t - B \sin\omega_\epsilon t\Big] 
- \Big[A\cos(\omega+\omega_\epsilon)t + B \sin(\omega+\omega_\epsilon)t
\Big] \Bigg\} \, ,
\label{fid1_ap}
\end{eqnarray}
\end{widetext}

\n where
\begin{equation}
\omega_\epsilon = {\epsilon(V_{22}-V_{11}) \over\hbar} \;\; ; \;\;
A+iB = \langle\alpha|\phi_1\rangle\langle\phi_2|\alpha\rangle V_{12} \, .
\end{equation}

\n Thus the first order approximation to the fidelity is the sum of three
periodic functions each of them oscillating with frequencies $\omega$,
$\omega_\epsilon$ and $(\omega + \omega_\epsilon)$. Then the periodicity
condition is
\begin{equation}
{\omega_\epsilon\over\omega} = 
{\epsilon (V_{22}-V_{11}) \over E_2-E_1} = {r\over s} \, ,
\end{equation}

\n with $r,s$ being co-primes. If $\omega_\epsilon=\omega$, the period of
oscillation in this approximation 
\begin{equation}
T' = 2\pi\hbar/|E_2-E_1| \, . 
\end{equation}

\n For $\omega_\epsilon\neq\omega$, we have
\begin{equation}
r(r+s)\omega = s(r+s)\omega_\epsilon = rs(\omega_\epsilon+\omega)
\end{equation}

\n and hence 
\begin{equation}
T' = 2\pi\hbar/|r(r+s)(E_2-E_1)| \, .
\end{equation}

\n That is, if the difference between first order energy corrections is
a rational multiple of the unperturbed energy level spacing, the quantum
fidelity is approximately periodic with period $T'$. In other words,
by reversing the perturbed dynamics at $t=jT'$ the initial state
$|\alpha\rangle$ is {\it nearly} recovered. 

If $V_{11} = V_{22}$ ($\omega_\epsilon = 0$), eq. (\ref{fid1_ap})
reduces to $f(t)\approx 1$. A trivial example of this case is that
$V=\hbox{constant}$, corresponds to the shifting of levels which is
dynamically insignificant. If the initial state is such that
$|u_1| = |u_2|^2$, then the first order approximation reduces to 
\begin{equation}
f(t) \approx {1\over 2} \Big[1 + \cos\omega_\epsilon t\Big]
\end{equation}

\n with $T'=2\pi\hbar/|\epsilon(V_{22}-V_{11})|$.

\section{Time dependent case}

Having identified the scenario of periodic echo in time independent two-level
systems, here we dwell on such a possibility for time dependent problems.
Considering the eigenvalue equation of $H$, as given in eq. (\ref{eigen}),
the initial state is rewritten as
\begin{equation}
|\alpha\rangle = c_1(0)|\phi_1\rangle + c_2(0)|\phi_2\rangle \, ,
\end{equation}

\n with the normalization, $|c_1(0)|^2 + |c_2(0)|^2 = 1$. Taking
$\tilde H = H + \epsilon V(t)$, the required evolutions are
\begin{equation}
\begin{array}{lll}
U(t)|\alpha\rangle &=& 
\sum_n c_n(0) e^{-iE_nt/\hbar} |\phi_n\rangle\, ,\\[4pt]
\tilde U(t)|\alpha\rangle &=& 
\sum_m c_m(t) e^{-iE_mt/\hbar} |\phi_m\rangle \, ,
\end{array}
\label{evolve}
\end{equation}

\n with $c_m(t)$ satisfying the Schr\"{o}dinger equation
\begin{equation}
{i\hbar\over\epsilon} \left[\begin{array}{c} \dot{c}_1(t) \\[4pt] 
\dot{c}_2(t) \end{array}\right] =
\left[\begin{array}{ll} V_{11}(t) & V_{12}(t)\;e^{-i\omega t} \\[4pt] 
V^{*}_{12}(t)\;e^{i\omega t} & V_{22}(t) \end{array}\right]
\left[\begin{array}{c} c_1(t) \\[4pt] c_2(t) \end{array}\right] \, ,
\label{sch}
\end{equation}

\n where $V_{mn}(t) = \langle\phi_m|V(t)|\phi_n\rangle$ and
$\omega = (E_2 - E_1)/\hbar$. We shall note that the solution of the above
system of differential equations is exactly known only for a few specific
forms of perturbation (see for example, ref. \cite{solved}).
Using eq. (\ref{evolve}) in (\ref{fid}), the quantum fidelity can be now
expressed as 
\begin{eqnarray}
f(t) &=& |c_1(t)|^2|c_1(0)|^2 + |c_2(t)|^2|c_2(0)|^2 \nonumber \\
&& \hspace*{1cm}
+ \;2\;\hbox{Re}\Big\{c^{*}_1(t)c_2(t)c_1(0)c^{*}_2(0)\Big\} \, .
\label{fid2}
\end{eqnarray}

\n That is, if the transition probabilities $|c_1(t)|^2$ and $|c_2(t)|^2$
are periodic in time the system could possibly exhibit periodic echo.
A simplest such situation is an oscillating perturbation of the form
$V_{11}(t) = V_{22}(t) = \mu$ and $V_{12}(t) = e^{i\nu t}$, where $\nu$ is
the perturbation frequency. 

Defining $\delta = \nu - \omega$, for the above choice of $V(t)$ the
Schr\"{o}dinger equation (\ref{sch}) may be rewritten as
\begin{equation}
\ddot{c}_1 - i\left[ \delta - {2\epsilon\mu\over\hbar} \right] 
\dot{c}_1 + {1\over \hbar^2} \left[ \epsilon^2(1-\mu^2) + \epsilon\mu\hbar 
\delta \right]c_1 = 0\, ,
\label{dif_c1}
\end{equation}

\n and a similar equation for $c_2(t)$ with $\delta$ being replaced by
$-\delta$. Defining the variables 
\begin{equation}
\Omega_r = {2\epsilon\over\hbar} \;\; ; \;\;
\Omega = \sqrt{\delta^2 + \Omega_r^2} \, ,
\end{equation}

\n the general solution to eq. (\ref{dif_c1}) is
\begin{eqnarray}
c_1(t) &=& \exp\left[ it\Big( {\delta\over 2} - {\epsilon\mu\over\hbar}
\Big) \right] \left\{ c_1(0)\cos\Big({\Omega t\over 2}\Big) \right. 
\hspace*{1.7cm} \nonumber \\
&& \hspace*{0.8cm} - \left. {i\over\Omega} \; \Big[\delta c_1(0)
+ \Omega_r c_2(0)\Big] \sin\Big({\Omega t\over 2}\Big)  \right\} \, .
\label{solu_c1}
\end{eqnarray}

\n To deduce the above solution, we note that with the choice  
\begin{equation}
c_1(t) = e^{i\eta t} x(t), \;\; 
\eta = {\delta\over 2} - {\epsilon\mu\over\hbar} \, ,
\label{form_c1}
\end{equation}

\n eq. (\ref{dif_c1}) is reduced to the harmonic oscillator equation,
$\ddot{x} + (\Omega/2)^2x = 0$. From the general solution of the latter,
after using (\ref{form_c1}) and (\ref{sch}), one obtains the solution
(\ref{solu_c1}). We also note that in the above example, resonance
condition corresponds to $\nu=\omega$, which implies $\delta = 0$ or
$\Omega=\Omega_r$. Proceeding further with either the substitution of
solution (\ref{solu_c1}) in eq. (\ref{sch}) or solving $c_2(t)$ in a similar
way, we obtain 
\begin{eqnarray}
c_2(t) &=& \exp\left[ -it\Big( {\delta\over 2} + {\epsilon\mu\over\hbar}
\Big) \right] \left\{ c_2(0)\cos\Big({\Omega t\over 2}\Big) \right.
\hspace*{1.5cm} \nonumber \\
&& \hspace*{0.8cm} + \left. {i\over\Omega} \; \Big[\delta c_2(0)
- \Omega_r c_1(0)\Big] \sin\Big({\Omega t\over 2}\Big) \right\} \, .
\label{solu_c2}
\end{eqnarray}

In what follows, it is useful to define the quantities 
\begin{equation}
D = |c_2(0)|^2 - |c_1(0)|^2 \;\; ; \;\; a+ib = c_1(0)\,c^{*}_2(0) \, .
\end{equation} 

\n Using the above expressions for $c_1(t)$ and $c_2(t)$, we have the
following:
\begin{widetext}
\begin{eqnarray}
|c_1(t)|^2 &=& {1\over 2\Omega^2} \Big[(\delta^2+\Omega^2)|c_1(0)|^2 
+ \Omega_r^2|c_2(0)|^2 \Big] + {a\delta\Omega_r\over\Omega^2} 
- {1\over 2\Omega^2} \Big[D\Omega_r^2 + 2a\delta\Omega_r\Big] \cos(\Omega t) 
- {b\Omega_r\over\Omega}\sin(\Omega t) \, , \nonumber \\[6pt]
|c_2(t)|^2 &=& {1\over 2\Omega^2} \Big[(\delta^2+\Omega^2)|c_2(0)|^2 
+ \Omega_r^2|c_1(0)|^2 \Big] - {a\delta\Omega_r\over\Omega^2} 
+ {1\over 2\Omega^2} \Big[D\Omega_r^2 + 2a\delta\Omega_r\Big] \cos(\Omega t) 
+ {b\Omega_r\over\Omega}\sin(\Omega t) \, , 
\label{gen1}
\end{eqnarray}

\n and 
\begin{eqnarray}
2\;\hbox{Re}\Big\{c^{*}_1(t)c_2(t)c_1(0)c^{*}_2(0)\Big\}
&=& {1\over\Omega^2} \; \Bigg\{ \Omega_r (2a\Omega_r - \delta D)
\Big[a\cos(\delta t) + b\sin(\delta t)\Big] 
+ {1\over 2} \Big[A_-\sin(\delta+\Omega)t
+ B_-\cos(\delta+\Omega)t\Big] \nonumber \\
&& \hspace*{4cm} + {1\over 2} \Big[A_+\sin(\delta-\Omega)t
+ B_+\cos(\delta-\Omega)t\Big] \Bigg\} \, ,
\label{gen2}
\end{eqnarray}

\n where 
\[ A_{\pm} = bD\Omega_r(\delta\pm\Omega) - 2ab\Omega_r^2 \;\;\; ; \;\;\; 
B_{\pm} = aD\Omega_r(\delta\pm\Omega)
+ 2 \Big[ a^2\delta^2 + b^2\Omega^2 \pm \delta\Omega(a^2+b^2) \Big] \, . \]
\end{widetext}
\n It is now easy to verify the normalization condition, that is
$|c_1(t)|^2 + |c_2(t)|^2 = 1$, which is due to the unitary quantum
evolution. If the initial state is such that $c_1(0) = 1$ and $c_2(0) = 0$,
then $a=b=0$ and $D=-1$. For this choice of initial state, at resonance
($\delta = 0$ or $\Omega=\Omega_r$) we have 
\begin{equation}
|c_1(t)|^2 = {1\over 2} \Big[1 + \cos(\Omega_r t)\Big] \, .
\end{equation}

\n Therefore, at resonance if the system is initially in one of the
unperturbed levels, the probability to remain in the same level oscillates
periodically with the frequency $\Omega_r$. This is the well known Rabi
oscillation and $\Omega_r$ is the so called Rabi frequency.

From the eqs. (\ref{gen1}) and (\ref{gen2}) we observe that the fidelity is
the sum of four periodic functions with frequencies $\delta, \Omega,
(\delta-\Omega)$ and $(\delta + \Omega)$. At resonance, eq. (\ref{gen1}) 
becomes
\begin{eqnarray}
|c_1(t)|^2 &=& {1\over 2}\Big[1 - D\cos(\Omega_r t) 
- 2b\sin(\Omega_r t)\Big] \, , \nonumber \\
|c_2(t)|^2 &=& {1\over 2}\Big[1 + D\cos(\Omega_r t)
+ 2b\sin(\Omega_r t)\Big] \, , 
\end{eqnarray}

\n and (\ref{gen2}) is reduced to
\begin{eqnarray}
2\;\hbox{Re}\Big\{c^{*}_1(t)c_2(t)c_1(0)c^{*}_2(0)\Big\} \hspace*{3cm} 
\nonumber \\
= 2a^2 + 2b^2\cos(\Omega_r t) - bD\sin(\Omega_r t) \, .
\end{eqnarray}

\n With this, the quantum fidelity takes the simple form 
\begin{equation}
f(t) = {1\over 2} + 2a^2 + \left\{{D^2\over 2} 
+ 2b^2\right\}\cos(\Omega_r t) \, .
\label{fidt}
\end{equation}

\n This shows that at resonance the quantum fidelity oscillates sinusoidally 
between $1$ and $4a^2$ with Rabi frequency. In other words, periodic behaviour
of fidelity with period $T=2\pi/|\Omega_r|=\pi\hbar/|\epsilon|$ is indeed a
direct consequence of the Rabi oscillation seen in transition probabilities.
Note that here $\epsilon$ is arbitrary and the above calculations are exact.
If $\epsilon$ is small the period $T$ is large. For integer $j$, it is now
easy to check that 
\begin{equation}
f(jT) = {1\over 2} + {1\over 2} \Big\{|c_1(0)|^2+|c_2(0)|^2\Big\}^2 = 1 \, .
\end{equation}

\n That is, upon inverting the perturbed evolution at $t=jT$, the initial
state is completely recovered. We may also consider this result as a
simple mechanism to generate periodic quantum echo for a two-level system.

It should be emphasized that periodic perturbation does not necessarily
imply that fidelity is periodic for arbitrary initial state. As it is
evident from the eq. (\ref{fidt}), $f(t)$ depends also on the choice of
initial state. If the initial state is an eigen state of the unperturbed
system i.e., $|\alpha\rangle = |\phi_1\rangle$ or $|\phi_2\rangle$,
$a=b=0$ and $D=\mp 1$. In this case, eq. (\ref{fidt}) becomes
\begin{equation}
f(t) = {1\over 2} \Big[1+\cos(\Omega_rt)\Big]
\end{equation}

\n and the fidelity oscillates between 1 and 0. On the other hand, if
$|\alpha\rangle = (|\phi_1\rangle \pm |\phi_2\rangle)/\sqrt{2}$ then
$f(t)=1$ and the fidelity is invariant in time. In other words, unperturbed
and perturbed evolutions are one and the same. For all other choices of
initial states, quantum dynamics is between the above two extremes. 

\section{Conclusions}

We have investigated the possibility of periodic dynamical echo for isolated 
two-level systems by computing the quantum fidelity. Considering two time
independent Hamiltonians, the fidelity is shown to be quasiperiodic in
general. It can be made periodic provided the ratio of corresponding
eigen-energy spacings are rational. In terms of perturbative approximation,
fidelity is nearly periodic if the difference between first order energy
corrections is a rational multiple of unperturbed energy level spacing.
The periodicity in fidelity implies that the two-level systems can exhibit
periodic quantum echo. Further, we have considered a specific time-dependent
problem, that is, a two-level system with oscillating time dependent
perturbation. If the perturbation frequency is in resonance with atomic
frequency, the fidelity is shown to display Rabi oscillation with period of
oscillation inversely proportional to the strength of the perturbation.
This is identified as a simple mechanism for generating periodic quantum
echo. 

It is yet to be seen if such periodic echoes of qubit can be exploited for
any useful information processing. In this context, it will be interesting
to extend similar studies for other exactly solvable problems and also for
multilevel systems. \\

\n {\bf Acknowledgement} \\
The work reported here forms part of the Council of Scientific and Industrial
Research and the Department of Science \& Technology, Government of India
supported research projects.

\end{document}